\documentclass[pdftex,twocolumn,epjc3]{svjour3}          

\RequirePackage[T1]{fontenc}

\smartqed  

\RequirePackage{graphicx}
\RequirePackage{mathptmx}      
\RequirePackage{flushend}
\RequirePackage[numbers,sort&compress]{natbib}
\RequirePackage[colorlinks,citecolor=blue,urlcolor=blue,linkcolor=blue]{hyperref}

\usepackage{amssymb}
\usepackage{amsmath}
\usepackage{epstopdf}
\RequirePackage{float}
\usepackage{bbm}
\usepackage{mathrsfs}
\usepackage{xcolor}

\newcommand{\phiout}{\phi^\text{out}}
\newcommand{\phir}{\phi^\text{rg}}

\newcommand{\curl}{\boldsymbol{\nabla} \times}
\newcommand{\vecx}{\mathbf{x}}

\journalname{Eur. Phys. J. C}

\begin{document}
\title{Repulsive Casimir-Lifshitz pressure in closed cavities
}
\author{C. Romaniega\thanksref{e1,addr1}
}
\thankstext{e1}{e-mail: cesar.romaniega@uva.es}
\institute{Departamento de F\'{\i}sica Te\'orica, At\'omica, y \'Optica, Universidad de Valladolid, 47011 Valladolid, Spain.\label{addr1}
}
\date{Received: date / Accepted: date}
\maketitle
\begin{abstract}
We consider the interaction  pressure acting on the surface of a dielectric sphere enclosed within a magnetodielectric cavity.  We determine the sign of this quantity regardless of the geometry of the cavity for systems at thermal equilibrium, extending the Dzyaloshinskii-Lifshitz-Pitaevskii result for homogeneous  slabs. 
As in previous theorems regarding Casimir-Lifshitz forces, the result is based on the scattering formalism. In this case the proof follows from the variable phase approach of electromagnetic scattering. 
With this, we present configurations in which both the interaction and the self-energy contribution to the pressure tend to expand the sphere.
\end{abstract}

\section{Introduction}\label{sec:I}
The Casimir effect, as one of the major macroscopic manifestations of quantum field theory, plays a fundamental role in micrometer and nanometer scale physics.
The  experimental accessibility, together with the possibility of technological
applications, requires a comprehensive knowledge  of this phenomena.
Although this is the case for simple configurations \cite{bordag2009advances,milton2001casimir}, we lack general theorems regarding the strong dependence of the force on  geometry and boundaries. 
For instance,  whether the force between two arbitrary bodies is attractive or repulsive is in general not known until the explicit calculation is performed. Only for a mirror symmetric arrangement of objects it has been  proved to be  attractive \cite{kenneth2006opposites,bachas2007comment}. As a result, for a single object in front of a plane the force is  attractive when both share  boundary conditions.   This may lead to a common cause of malfunction of nanoscale and microscale  machines: the permanent adhesion of their moving parts, known as stiction \cite{buks2001stiction,munday2010repulsive}. In this sense, different methods for obtaining repulsive forces have been proposed. Back in 1974, 
dielectric-magnetic systems  were introduced by Boyer \cite{boyer1974van}, the use of metamaterials \cite{rosa2008casimir,zhao2009repulsive} or topological insulators \cite{grushin2011tunable,rodriguez2014repulsive} has been discussed lately, as well as configurations with nontrivial geometry \cite{levin2010casimir} and nontrivial topology \cite{abrantes2018repulsive}. Other proposals are not based on  particular parameters or shapes of  
materials, which might make experimental realization challenging, but on the introduction of an intermediate medium.
This was the first prediction of a repulsive interaction between two objects, developed by Dzyaloshinskii-Lifshitz-Pitaevskii (DLP) in 1961  \cite{dzyaloshinskii1961general}. They considered two parallel homogeneous slabs  separated by another material with nontrivial  electromagnetic response.
The force across the medium was found to be proportional to
\begin{equation}\label{eq:LifshitzRel}
	-(\varepsilon_1-\varepsilon_{M})(\varepsilon_2-\varepsilon_{M}),
\end{equation}
This behavior, hereinafter referred to as the DLP result, leads to repulsion if the permittivities of the objects $\varepsilon_{i}$ and the medium $\varepsilon_{M}$ satisfy  $\varepsilon_{i}<\varepsilon_{M}<\varepsilon_{j}$.
This also resulted in the first experimental confirmation of a repulsive interaction between material bodies: a gold-covered sphere
and a large silica plate immersed in bromobenzene  \cite{munday2009measured}.

In addition to the sign of the force,  its magnitude \cite{venkataram2020fundamental} and the stability should be considered for the design
of  mechanical and levitating devices, in particular when looking for ultra-low stiction \cite{capasso2007casimir,munday2009measured}. In this context,
an extension of Earnshaw's
theorem sets restrictive constraints on the stability of neutral objects held in equilibrium by Casimir-Lifshitz forces \cite{rahi2010constraints}. The result is based on the scattering approach, an analysis similar to the one performed in \cite{kenneth2006opposites}.  For instance, in the presence of two nonmagnetic bodies these  conditions are completely determined by the sign of  expression \eqref{eq:LifshitzRel}, excluding stable
equilibria when the objects are immersed in vacuum.  However, it is worth noting that
the introduction of a chiral medium  could avoid the assumptions of the previous two no-go theorems \cite{kenneth2006opposites,rahi2010constraints}, leading to measurable forces varying in response to an external magnetic field \cite{jiang2019chiral}.

\begin{figure}[h!]
	\centering
	\includegraphics[width=0.48\textwidth]{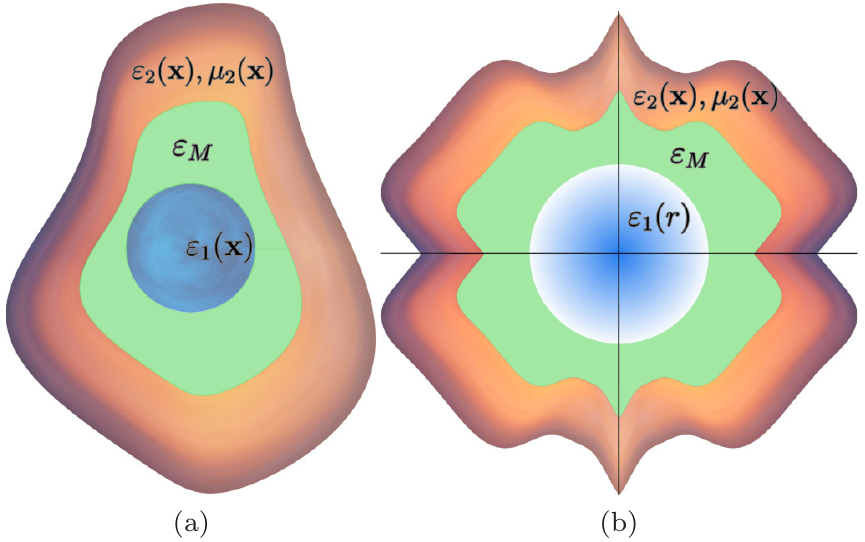}
	\caption{Sketch of the system under study. In the configuration  on the right the sum of the Casimir-Lifshitz forces on each object equals zero. In both cases the    force acting on the surface of the sphere due to the interaction with the magnetodielectric cavity can be considered. Between both objects there is another frequency-dependent, but homogeneous, dielectric.}\label{fig:conf}
\end{figure}

The above-mentioned work has  focused primarily on configurations in which the bodies lie outside each other, even though closed cavities are experimentally realizable \cite{marachevsky2001casimir,hoye2001casimir,brevik2002casimir,brevik2005casimir,dalvit2006exact,marachevsky2007casimir, zaheer2010casimir,teo2010casimir,rahi2010stable,parashar2017electromagnetic}.
In Sec.~\ref{sec:II} we  study such configurations within the scattering framework \cite{kenneth2008casimir,rahi2009scattering}.
The main result of the text is presented in Sec.~\ref{sec:III}, stating that the sign of the pressure acting on the surface of an inhomogeneous dielectric sphere due to the interaction with  an arbitrarily shaped cavity is completely determined by the sign of expression \eqref{eq:LifshitzRel}. 
The derivation is based on a simple result of electromagnetic scattering and it is easily extended to magnetodielectric cavities and systems at thermal equilibrium.
In Sec.~\ref{sec:TotalPressure} we include the self-energy contribution to the pressure for a dilute dielectric ball. As a consistency test, we also recover the DLP result, and its extension to inhomogeneous slabs, showing the relation between the interaction pressure on the sphere and the force between the slabs. We end in Sec.~\ref{sec:IV} with some remarks on the main result and the conclusions. 
\\
{\section{Interaction energy}\label{sec:II}}
In cavity configurations invariant under mirror symmetry  with respect to the three spatial planes [Fig.~\ref{fig:conf}(b)],  the sum of the Casimir-Lifshitz forces  on each object equals zero. 
However, the pressure acting on their surfaces does not vanish. As already noted, we will focus on this quantity for a sphere inside a cavity [Fig.~\ref{fig:conf}(a)]. 	Using the $TGTG$ representation of the interaction energy we will  determine the sign of this pressure as a function of the permittivities and permeabilities of the bodies and the medium.  Indeed, we will see that the results derived remain valid when the sphere is outside the cavity. 
Throughout this paper we will use the natural units  $\hslash=c=\varepsilon_0=\mu_0=1$, neglecting fluctuations 
due to nonelectromagnetic oscillations when the medium is different from vacuum, which are usually small \cite{dzyaloshinskii1961general}.   

We assume that the coupling of the  electromagnetic field to matter can be described by  continuous permittivity $\varepsilon$ and permeability ${\mu}$ functions. For a
homogeneous medium characterized by  $\varepsilon_{M}$ and $\mu_{M}$, Maxwell curl equations, 
after Fourier transform in time, can be rearranged  to give a stationary vector Schr\"odinger-like equation \cite{newton2013scattering}
\begin{equation}\label{eq:SchEM}
	\left[ \boldsymbol{\nabla} \times \boldsymbol{\nabla} + \mathbb{V}(\omega,\mathbf{x})\right]\textbf{E}(\omega,\mathbf{x}) =k^2
	\textbf{E}(\omega,\mathbf{x}),
\end{equation}
where $k\equiv\sqrt{\varepsilon_{M}}\omega$ and the potential  operator $\mathbb{V}(\omega,\mathbf{x})$ is
\begin{equation*}	
	\mathbb{I} {\omega^2} \left[\varepsilon_{M}(\omega)-\varepsilon(\omega, \textbf{x})\right] +\boldsymbol{\nabla} \times \left[\frac{1}{\mu(\omega,\textbf{x})}-\frac{1}{\mu_{M}(\omega)}\right] \boldsymbol{\nabla} \times.
\end{equation*}
Since the magnetic response of ordinary materials is typically close to one, we will focus on nonmagnetic bodies.  However, we shall see that in some cases the introduction of nontrivial permeabilities  poses no additional difficulties, especially for the cavity. 
In any case, we deal with two nonoverlapping bodies. Specifically, if ${\mu(\omega,\textbf{x})}={\mu_{M}(\omega)}=1$ we have
\begin{equation}\label{eq:Vi}
	\mathbb{V}_{i}(\omega, \textbf{x}) = \mathbb{I} {V}_{i}(\omega, \textbf{x}) =\mathbb{I} \omega^2 \left[\varepsilon_{M}(\omega)-\varepsilon_{i}(\omega, \textbf{x})\right],
\end{equation}
being $\text{supp}\,V_1\cap\,\text{supp}\,V_2=\emptyset$. Here $\text{supp}$ stands for the spatial support of the function. Consequently, the Casimir interaction energy between the two objects is encapsulated in the so-called \textit{TGTG} formula \cite{kenneth2008casimir,rahi2009scattering}
\begin{equation}\label{eq:ETGTG}
	E_\text{int}=\frac{1}{2\pi}\int_0^\infty d\kappa\,\text{Tr} \log(\mathbb{I}- {\mathbb{T}_1}\mathbb{G}^{M}_{12}\mathbb{{T}}_2\mathbb{G}^{M}_{21}).
\end{equation}
As usual, we  will carry out the integration over imaginary frequencies $\omega= i \kappa$.  In this sense, we denote by $\varepsilon(i\kappa)$ the analytic continuation of the permittivity to the imaginary frequency axis, which from Kramers-Kronig causality conditions satisfies $\varepsilon(i\kappa)\geq 1$ \cite{bordag2009advances}.
The properties of each body are encoded in the Lippmann-Schwinger $\mathbb{T}$ operator  $\mathbb{T}_{i}:\mathcal{H}_{i}\to \mathcal{H}_{i}$, being $\mathcal{H}_{i}\equiv L^2(\text{supp}\,\mathbb{V}_{i})^3$ \cite{hanson2013operator}. Furthermore, the relative position between both objects  enters through the operator
$
\mathbb{G}^{M}_{ij}\equiv \mathbb{P}_{i}\mathbb{G}^{M} \mathbb{P}_{j}:\mathcal{H}_{j}\to \mathcal{H}_{i},
$
being $\mathbb{G}^{M}$  the  propagator across the medium and $\mathbb{P}_{i}$ the projection operator onto the Hilbert space  $\mathcal{H}_{i}$.
The electric Green's dyadics  fulfill
\begin{equation*}
	\left[ \boldsymbol{\nabla}\times\boldsymbol{\nabla}\times+\varepsilon_{M}(i \kappa){\kappa^2} \right] \mathbb{G}^{M}(i\kappa,\mathbf{x},\mathbf{x}') =~ \mathbb{I}\delta (\mathbf{x}~ -~ \mathbf{x}'),
\end{equation*}
which are related to the vacuum functions by 
\begin{equation*}
\mathbb{G}^0(i \sqrt{\varepsilon_{M}}\kappa,\mathbf{x},\mathbf{x}') = \mathbb{G}^{M}(i \kappa,\mathbf{x},\mathbf{x}').
\end{equation*}
Writing $\mathbb{G}^{M}$ in terms of the Green's function for the scalar Helmholtz equation, it can be proved that 
$\langle\textbf{E}, \mathbb{G}^{M}\textbf{E} \rangle\geq~0
$
for all the vectors $\textbf{E}$ \cite{hanson2013operator,rahi2010constraints}.
Namely, $ \mathbb{G}^{M} $ is a nonnegative operator  $\mathbb{G}^{M}\geq~0$. 
In addition, these functions are related to the $\mathbb{T}_{i}$ operators by
$
\mathbb{T}_{i} = \mathbb{V}_{i} /({\mathbb{I} + \mathbb{G}^{M} \mathbb{V}_{i}}),
$
where the Lippmann-Schwinger equation of electromagnetic scattering is formally written as
$ 
\textbf{E}  = \textbf{E}_0- \mathbb{G}^{M} \mathbb{T}_i \textbf{E}_0.
$
We will not analyze convergence issues or the  self-adjointness of the presented operators (for those see \cite{hanson2013operator}), assuming that the appropriate  conditions are fulfilled in realistic systems \cite{rahi2010constraints}.

Since the two bodies are  separated from each other,  we can expand the Green's functions in terms of free solutions  of Eq.~\eqref{eq:SchEM}. In spherical coordinates there is a regular solution at the origin $\textbf{E}_{a}^\text{rg}(\omega, \textbf{x})$, whose radial part is determined by the spherical Bessel function $j_\ell (\omega r)$, and an outgoing (incoming) solution   $\textbf{E}_{a}^\text{out}(\omega, \textbf{x})$ [$\textbf{E}_{a}^\text{in}(\omega, \textbf{x})$], whose radial part is determined by the spherical Hankel function of the first (second) kind $h^{(1)}_\ell (\omega r)$ [$h^{(2)}_\ell (\omega r)$] \cite{sun2019invariant}. The subscript of these transverse solutions stands for the angular momentum values $\{\ell, {m}\}$ and  polarizations. Based on the appropriate representation of $\mathbb{G}^{M}_{ij}$ when the two bodies lie entirely outside each other, the operator  ${\mathbb{T}_1}\mathbb{G}^{M}_{12}\mathbb{{T}}_2\mathbb{G}^{M}_{21}$  in Eq.~\eqref{eq:ETGTG} can be expanded as
\begin{equation}\label{eq:ExtConf}
	  \langle\textbf{E}_{j}^\text{rg}, {\mathbb{T}_1} \textbf{E}_{a}^\text{rg} \rangle \langle  \textbf{E}_{a}^\text{rg}, \mathbb{G}^{M}_{12} \textbf{E}_{b}^\text{rg} \rangle \langle  \textbf{E}_{b}^\text{rg}, \mathbb{{T}}_2 \textbf{E}_{c}^\text{rg} \rangle \langle  \textbf{E}_{c}^\text{rg}, \mathbb{G}^{M}_{21} \textbf{E}_k^\text{rg} \rangle,
\end{equation}
where a sum over $a, b, c$ is assumed.
In this case $\langle\textbf{E}_{j}^\text{rg},  {\mathbb{T}_{i}} \textbf{E}_{k}^\text{rg}\rangle $ encodes the usual scattering amplitude  related to  a  process in which a regular wave
interacts with an object and scatters outward \cite{newton2013scattering}.
The  sum changes to
\begin{equation}\label{eq:IntConf}
	   \langle  \textbf{E}_{j}^\text{rg}, {\mathbb{T}_1} \textbf{E}_{a}^\text{rg} \rangle \langle  \textbf{E}_{a}^\text{rg}, \mathbb{G}^{M}_{12} \textbf{E}_{b}^\text{in} \rangle \langle  \textbf{E}_{b}^\text{in}, \mathbb{{T}}_2 \textbf{E}_{c}^\text{out} \rangle \langle  \textbf{E}_{c}^\text{out} , \mathbb{G}^{M}_{21} \textbf{E}_k^\text{rg} \rangle 
\end{equation}
for interior configurations, where one body is inside the other. In this less common version of the $TGTG$ representation, the usual scattering amplitude  arises for the first body and  $\langle  \textbf{E}_{b}^\text{in}, \mathbb{{T}}_2 \textbf{E}_{c}^\text{out} \rangle$ for the cavity. 
The latter is associated with a scattering experiment in which the source and the detector are inside the cavity \cite{rahi2009scattering}. In this regard, expansion \eqref{eq:IntConf} offers a schematic description of the travel of the wave between both bodies: a regular wave reaches the first body, part of the wave is reflected as an outgoing wave heading towards the cavity, where is partially scattered as an incoming wave, contributing to form the regular wave which reaches the first body and the process is repeated. This iteration clearly reveals the nonadditive character of fluctuation-induced forces.  In line with this, an alternative proof of the $TGTG$ formula for symmetric bodies based on the mode summation approach is given in \cite{teo2012mode}.  
Although the $TGTG$ representation in terms of operators is formally the same, we want to emphasize that the suitable Green's function expansions depend on the configuration.

In order to determine the sign of the Casimir energy we assume that the sign $s_{i}$ of the potential $V_{i}$ in Eq.~\eqref{eq:Vi} is  constant over the whole body, being
\begin{equation}\label{eq:Signs}
	s_{i}=\pm 1 \ \ \text{if} \ \ \varepsilon_{i}(i\kappa, \textbf{x}) \gtrless \varepsilon_{M}(i \kappa), \ \ \forall \textbf{x}\in \text{supp}\,V_{i}.
\end{equation}
In this case $\mathbb{T}_{i}(i\kappa)$  is real and symmetric and can be written in the form 
$
\mathbb{T}_{i}=s_{i} \sqrt{s_{i}\mathbb{T}_{i}} \sqrt{s_{i}\mathbb{T}_{i}},
$
where $\sqrt{s_{i}\mathbb{T}_{i}}$ is the square root of the positive operator $s_{i} \mathbb{T}_{i}$ \cite{rahi2010constraints}.
We shall see that our analysis applies to each fixed frequency so Eq.~\eqref{eq:Signs} should hold for all of them. However,  we can simply assume constant sign over the frequencies  contributing most to the energy \cite{dzyaloshinskii1961general,munday2009measured}.
Accordingly, the interaction energy can be rewritten as
\begin{equation}\label{eq:EnM}
	E_\text{int}=\frac{1}{2\pi}\int_0^\infty d\kappa\,\text{Tr} \log(\mathbb{I}-s\, \mathbb{M}),
\end{equation}
where we have defined
$
\mathbb{M} \equiv\sqrt{s_2\mathbb{T}_2}\mathbb{G}^{M}_{21} {s_1\mathbb{T}_1} \mathbb{G}^{M}_{12} \sqrt{s_2\mathbb{T}_2}
$
and $s\equiv s_1 s_2$.
The representations \eqref{eq:ETGTG} and \eqref{eq:EnM} are equivalent since
$$
\text{Tr} \log(\mathbb{I}-s\, \mathbb{M}) = \text{Tr} \log(\mathbb{I}-{\mathbb{T}_1}\mathbb{G}^{M}_{12}\mathbb{{T}}_2\mathbb{G}^{M}_{21}).
$$
This follows from $\text{Tr} \log(\mathbb{I}-s\, \mathbb{M})= \log \det (\mathbb{I}-s\, \mathbb{M})$ and the determinant identity $\det(\mathbb{I}-\mathbb{A}\mathbb{B})=\det(\mathbb{I}-\mathbb{B} \mathbb{A})$, where  $\mathbb{A}\mathbb{B}=s\, \mathbb{M}$ and $\mathbb{B}\mathbb{A}={\mathbb{T}_1}\mathbb{G}^{M}_{12}\mathbb{{T}}_2\mathbb{G}^{M}_{21}$ \cite{simon1977notes}. In addition, using $(\mathbb{G}^{M}_{ij})^\dagger=\mathbb{G}^{M}_{ji}$, we rewrite this new operator as
\begin{equation} \label{eq:Proof}
	\mathbb{M}=(\sqrt{s_1\mathbb{T}_1}\mathbb{G}^{M}_{12} \sqrt{s_2\mathbb{T}_2})^\dagger \sqrt{s_1\mathbb{T}_1}\mathbb{G}^{M}_{12} \sqrt{s_2\mathbb{T}_2},
\end{equation}
proving that it  is  nonnegative.
We will frequently  make use of this standard reasoning, which follows from the definition of the adjoint:
\begin{equation}\label{eq:Positive_Op}
	\langle  \textbf{E}, {\mathbb{C}^\dagger} \mathbb{C} \textbf{E} \rangle = \langle {\mathbb{C}} \textbf{E},  \mathbb{C} \textbf{E} \rangle =\Vert\mathbb{C} \textbf{E}\Vert^2\geq 0. 
\end{equation}
The eigenvalues of $\mathbb{M}$, besides being nonnegative, belong to $[0, 1)$.  This   has been proved for the operator $\mathbb{T}_1\mathbb{G}^0_{12}\mathbb{{T}}_2\mathbb{G}^0_{21}$ when the intermediate medium is vacuum,   $\mathbb{T}_{i}>0$, thus obtaining $E_\textup{int}~\leq~0$ \cite{kenneth2008casimir}. In our case, the same derivation holds replacing $\mathbb{T}_{i}$ and $\mathbb{G}^0$ by $s_{i}\mathbb{{T}}_{i}>0$ and $\mathbb{G}^{M}$, noting that the nonzero eigenvalues of $s \mathbb{A}\mathbb{B}= \mathbb{M}$ and  $s\mathbb{B}\mathbb{A}=s{\mathbb{T}_1}\mathbb{G}^{M}_{12}\mathbb{{T}}_2\mathbb{G}^{M}_{21}$ are the same. It is then clear that we can also obtain $E_\textup{int}~>~0$. First, using  Lidskii's theorem the trace in Eq.~\eqref{eq:EnM} can be expressed    as
\begin{equation*}
	\text{Tr} \log(\mathbb{I}-s\, \mathbb{M}) = \sum_\alpha \log(1-s\lambda_\alpha),\quad \mathbb{M}\textbf{E}_\alpha=\lambda_\alpha\textbf{E}_\alpha.
\end{equation*}
Consequently, if $\lambda_\alpha\neq 0$ we have   $s=- \textup{sgn}\log(1-s\lambda_\alpha)$ and $\textup{sgn}\, E_\textup{int}=-s$. The latter can be written in terms of the permittivities as
\begin{equation}\label{eq:SignEn}
	\textup{sgn}\,E_\textup{int}=-s=- \textup{sgn}\left[(\varepsilon_{1}-\varepsilon_{M}\right)(\varepsilon_{2}-\varepsilon_{M})].
\end{equation}
For magnetodielectric objects characterized by $\varepsilon_{i}$ and $\mu_{i}$, the relation $\textup{sgn}\, E_\textup{int}=-s$ remains valid as long as the sign of the differential operator $\mathbb{V}_{i}(i\kappa,\mathbf{x})$ in Eq.~\eqref{eq:SchEM} is well-defined. The latter is determined by Eq.~\eqref{eq:Signs} if we include an additional condition:
\begin{equation*}
	s_{i}=\pm 1 \quad \text{if} \  \varepsilon_{i}(i\kappa, \textbf{x}) \gtrless \varepsilon_{M}(i \kappa)\ \text{and}\ \mu_{i}(i \kappa, \textbf{x})~\lesseqgtr~\mu_{M}(i \kappa),
\end{equation*}
for the whole body \cite{rahi2009scattering}.
It is also worth emphasizing that the result on the sign of the energy \eqref{eq:SignEn} is  valid for two arbitrary bodies, even when they lie outside each other.
\\
\section{Interaction pressure}\label{sec:III}
The Casimir-Lifshitz interaction pressure acting on the surface of the sphere can be obtained from the representation of the  energy given in Eq.~\eqref{eq:EnM}.  To define this pressure we do not need to include elastic deformations, we simply make use of the principle of virtual work. Accordingly, the mean value of the pressure due to a virtual variation of the radius of the sphere $r_{1}$  satisfies  \cite{barton2004casimir,li2019casimir}
\begin{equation}\label{eq:MeanPress}
	\langle p_\text{int} \rangle\equiv \dfrac{1}{4\pi}\int_{S^2} d\Omega\, p_\text{int}(r_{1},\Omega)= -\dfrac{1}{4\pi r_{1}^{2}}\frac{\partial  E_\text{int}}{\partial  r_{1}}.
\end{equation}
In particular, for a spherically symmetric system  the pressure is constant 
$
\langle p_\text{int} \rangle=p_\text{int}(r_{1}).
$
In order to evaluate the sign of $\partial_{r_{1}}  E_\text{int}$
we employ the variable phase approach of electromagnetic scattering \cite{johnson1988invariant}. This method 
is progressively reaching some importance in Casimir physics since it enables to compute efficiently  $\mathbb{T}_i$ for nonsymmetric objects \cite{forrow2012variable}. Specifically, in order to prove $\partial_{r_{1}}(s_1\mathbb{T}_1)>0$ we make use of
the quantum mechanical Calogero equation \cite{calogero1967variable} generalized to electromagnetic scattering by arbitrarily shaped objects \cite{johnson1988invariant,sun2019invariant}
\begin{equation}\label{eq:Caloguero}
	\frac{\partial {T}_1}{\partial {r_{1}}}= - i k ({J}_1+{H}_1{T}_1)^T{U}_1({J}_1+{H}_1{T}_1).
\end{equation}
The superscript $T$ denotes the transpose operation and ${T}_1$ is the $\mathbb{T}_1$ operator in the spherical wave basis, i.e., $({T}_1)_{{a}{b}}$ stands for  $\langle  \textbf{E}_{a}^\text{rg}, {\mathbb{T}_1} \textbf{E}_{b}^\text{rg} \rangle$  of expansion \eqref{eq:ExtConf} in terms of the complete set of regular solutions. Furthermore, the potential $V_{1}$ defined in Eq.~\eqref{eq:Vi} enters through
\begin{equation*}
	\langle  \textbf{E}_{\ell {m}}^\text{rg}, {U}_1   \textbf{E}_{\ell' {m}'}^\text{rg} \rangle=r^2 \int_{S^2}  d\Omega\, {Y}^\dagger_{\ell {m}} V_1(\omega,\textbf{x}){D}(\omega, \textbf{x}){Y}_{\ell'{m}'},
\end{equation*}
being the matrix ${Y}_{\ell {m}}$  composed of vector spherical harmonics and ${D}(\omega, \textbf{x})  =   \text{diag}(\varepsilon_M(\omega) /\varepsilon_1(\omega, \textbf{x}), 1,1)$ \cite{sun2019invariant}. For imaginary frequencies we have defined the two real matrices ${J}_1$ and ${H}_1$
\begin{eqnarray*}
	\langle  \textbf{E}_{\ell {m}}^\text{rg},{J}_1(i \xi)   \textbf{E}_{\ell' {m'}} ^\text{rg} \rangle
	 = \frac{1}{r_{1}} \left(
	\begin{array}{cc}
		i_\ell(\xi  r_{1}) & 0 \\
		0 & \partial_{r_{1}} i_\ell(\xi  r_{1}) \\
		0 & \dfrac{i_\ell(\xi  r_{1})}{r_{1}} \\
	\end{array}
	\right) \delta_{\ell  \ell'}\delta_{{m}{m}'},\\[0.5ex]
	\langle  \textbf{E}_{\ell {m}}^\text{rg},{H}_1(i \xi)   \textbf{E}_{\ell' {m'}} ^\text{rg} \rangle
	 = \frac{1}{r_{1}} \left(
	\begin{array}{cc}
		k_\ell(\xi  r_{1}) & 0 \\
		0 & \partial_{r_{1}} k_\ell(\xi  r_{1}) \\
		0 & \dfrac{k_\ell(\xi  r_{1})}{r_{1}} \\
	\end{array}
	\right) \delta_{\ell  \ell'}\delta_{{m}{m}'},
\end{eqnarray*}
in terms of $\xi\equiv  {\kappa}{\sqrt{\varepsilon_{M}}}$ and the  modified Bessel functions  
$$
i_\ell(z)\equiv \sqrt{{\pi }/2 z}\, I_{\ell+{1}/{2}}(z),\quad k_\ell(z)\equiv \sqrt{2/\pi z}\, K_{\ell+{1}/{2}}(z).
$$
To reach this expressions we have used the same expansion of the background Green's functions as in \cite{johnson1988invariant,sun2019invariant}
\begin{equation*}
	\mathbb{G}^M({\bf x},{\bf x}')= \xi \sum_{a}
	\left\{ \begin{array}{cc} {\bf
			E}_{a}^{\rm out}({\bf x}) \otimes {\bf
			E}_{a}^{{\rm reg *}}({\bf x}')  & \ {\rm if}
		\ \ r> r',\\[0.5ex]{\bf E}_{a}^{{\rm
				reg}}({\bf x}) \otimes  {\bf E}_{  a}^{{\rm
				out*}}({\bf x}')  & \ {\rm if} \ \ r<r'.
	\end{array}\right.
\end{equation*}
Each mode of the electric field is determined by the transverse electric (TE) and magnetic (TM) vectors  
\begin{eqnarray*}
	{\mathbf{M}}^\text{rg}_{\ell m}(i \xi,\vecx)    &=& \frac{1}{\sqrt{\ell(\ell+1)}} \
	\curl \phir_{\ell m}(i \xi,\vecx)\, \vecx, \\[0.5ex]
	\mathbf{M}^\text{out}_{\ell m}(i \xi,\vecx) &=& \frac{1}{\sqrt{\ell(\ell+1)}}\
	\curl \phiout_{\ell m}(i \xi,\vecx) \, \vecx, \\[0.5ex]
	\mathbf{N}^\text{rg}_{\ell m}(i \xi,\vecx)    &=& \frac{1}{\xi \sqrt{\ell(\ell+1)}}\
	\curl \curl \phir_{\ell m}(i \xi,\vecx) \, \vecx, \\[0.5ex]
	\mathbf{N}^\text{out}_{\ell m}(i \xi,\vecx) &=& \frac{1}{\xi \sqrt{\ell(\ell+1)}}\
	\curl \curl \phiout_{\ell m}(i \xi,\vecx) \, \vecx,
\end{eqnarray*}
which can be expressed in terms of the scalar Helmholtz equation solutions \cite{rahi2009scattering}
$$
\phir_{\ell m}(i \xi,\vecx)  = i_l(\xi r) Y^m_{\ell}(\Omega),\quad
\phiout_{\ell m}(i \xi,\vecx)= k_l(\xi r) Y^m_{\ell}(\Omega).
$$

Nevertheless, the relevant fact is the structure of the right-hand side of Eq.~\eqref{eq:Caloguero}, which naturally leads to the positivity of $s_1\partial_{r_{1}}{T}_1$. Along the imaginary frequency axis ${T}_1$ is real and symmetric, consequently, using Eq.~\eqref{eq:Positive_Op}, $s_1\partial_{r_{1}}{T}_1$ is positive if $s_1 {U}_1$ is, and, for the same reason, the latter is positive if $s_1 V_1(i\kappa,\textbf{x})>0$, which holds trivially assuming \eqref{eq:Signs}.

As a consistency test, we prove by explicit calculation that $s_1\partial_{r_{1}}{\mathbb{T}_1}$ is positive for a spherically symmetric  object [Fig.~\ref{fig:conf}(b)]. In this case we can make use of the standard Lorenz-Mie  theory \cite{johnson1999exact}. The problem is completely decoupled for the angular momentum values and the two polarizations.
Indeed, electromagnetic scattering
reduces to two independent scalar  problems, one for each polarization \cite{johnson1999exact}. For instance, the two radial potentials for  a homogeneous sphere  are
$$
V_{1}^{\textup{TE}}(\omega)=V_{1}^{\textup{TM}}(\omega)=\omega^2 \left[\varepsilon_{M}(\omega)-\varepsilon_{1}(\omega)\right],
$$
being 
$
V_{1}^{\textup{TE}}(\omega,r) = \omega^2 \left[\varepsilon_{M}(\omega)-\varepsilon_{1}(\omega,r)\right]
$
valid for any spherically symmetric object \cite{toni2014advances}.
Consequently, we can write
\begin{equation}
	\langle \textbf{E}^\text{rg}_{\ell {m}} ,\mathbb{T}\, \textbf{E}^\text{rg}_{\ell' {m}'}\rangle = \left(
	\begin{array}{cc}
		{T}^\textup{TE}_{\ell}  & 0 \\
		0 &{T}^\textup{TM}_{\ell}  \\
	\end{array}\right)\delta_{\ell  \ell'}\delta_{{m}{m}'},
\end{equation}
where the subscript in $\mathbb{T}_1$ and the $i \kappa$ dependence have been omitted for simplicity.
Rotating to  imaginary frequencies the derivatives of the Lorenz-Mie coefficients \cite{rahi2009scattering,johnson1999exact}, we obtain the following first-order nonlinear differential equations:
\begin{eqnarray*}
	s_1\partial_{r_{1}}{T}^\textup{TE}_{\ell}&=&s_1\left({\varepsilon_1({r_{1}})}-{\varepsilon_{M}}\right) \frac{a_1^2 \kappa  {r_{1}} }{2 \pi {\varepsilon_{M}}},
	\\
	s_1\partial_{r_{1}}{T}^\textup{TM}_{\ell}&=&s_1\left({\varepsilon_1({r_{1}})}-\varepsilon_{M}\right)\frac{a_2^2 \ell (\ell+1)+a_3^2 {\varepsilon_1({r_{1}})}/{\varepsilon_{M}}}{2 \pi  \kappa  {r_{1}}\, {\varepsilon_1({r_{1}})} }.
\end{eqnarray*}
Both derivatives are positive since the three parameters $a_{i}$ are real:
\begin{eqnarray*}
	a_1 &=& -2 {T}^\textup{TE}_{\ell {m}} \kappa  K_{\ell+{1}/{2}}(\kappa {r_{1}})+\pi  I_{\ell+{1}/{2}}(\kappa {r_{1}}),\\[0.5ex]
	a_2 &=&    \pi  I_{\ell+{1}/{2}}(\kappa {r_{1}})+2 {T}^\textup{TM}_{\ell {m}} \kappa  K_{\ell+{1}/{2}}(\kappa {r_{1}}),\\[0.5ex]
	a_3 &=&   2 \kappa {T}^\textup{TM}_{\ell {m}}  \left[  (\ell+1) K_{\ell+{1}/{2}}(\kappa {r_{1}})-\kappa  {r_{1}} K_{\ell+{3}/{2}}(\kappa {r_{1}})\right]\\[0.5ex]
	&+&\pi \kappa  {r_{1}} I_{\ell+{3}/{2}}(\kappa {r_{1}})+\pi  (\ell+1) I_{\ell+{1}/{2}}(\kappa {r_{1}}).
\end{eqnarray*}

Having proved $\partial_{r_{1}}({s_1\mathbb{T}_1})>0$, we can straightforwardly find the sign of $\langle p_\textup{int}\rangle$. We simply note that only $\mathbb{T}_1$ depends on $r_{1}$  in $\mathbb{M}$, i.e.,
we can write
$$
\partial_{r_{1}} \mathbb{M}=\left(\mathbb{G}^{M}_{12} \sqrt{s_2\mathbb{T}_2}\right)^\dagger \partial_{r_{1}}({s_1\mathbb{T}_1}) \left(\mathbb{G}^{M}_{12} \sqrt{s_2\mathbb{T}_2}\right)>0.
$$
As before, the positivity is proved using Eq.~\eqref{eq:Positive_Op}. Applying the Hellmann-Feynman theorem to the eigenvalues of $\mathbb{M}$,  we obtain  $\partial_{r_{1}} \lambda_\alpha>0$ \cite{feynman1939forces}.
With this, 
$\textup{sgn}\,\partial_{r_{1}} \log(1-s\lambda_\alpha)=-s$. Finally, from Eq.~\eqref{eq:EnM} and Lidskii's theorem we obtain $\text{sgn}\,\partial_{r_{1}}  E_\text{int}=-s$, which can be written in terms of the pressure with Eq.~\eqref{eq:MeanPress} 
\begin{equation}\label{eq:SignPint}
	\textup{sgn}\langle p_\textup{int}\rangle= s = \textup{sgn}\left[(\varepsilon_{1}-\varepsilon_{M}\right)(\varepsilon_{2}-\varepsilon_{M})].
\end{equation}
This is the main result of our work. Since we have considered permittivity functions $\varepsilon(i\kappa, \textbf{x})$ such that the sign of $\varepsilon_{i}(i\kappa, \textbf{x})-\varepsilon_{M}(i\kappa)$ is independent of $\kappa$ and $\textbf{x}$, we have written $\textup{sgn}(\varepsilon_{i}-\varepsilon_{M})=\textup{sgn}[\varepsilon_{i}(i\kappa, \textbf{x})-\varepsilon_{M}(i\kappa)]$.

We now compare this result with particular  configurations previously studied in the literature. First, for two concentric spherical shells  satisfying perfectly conducting boundary conditions a positive pressure is obtained using the zeta function regularization method \cite{teo2010casimir}. This is consistent with Eq.~\eqref{eq:SignPint} since these idealized conditions arise in the limit of large permittivities. This positive pressure is also found in the  experimental setup suggested in \cite{brevik2005casimir}, where the same boundary conditions are considered using Green's functions.
Secondly, based on a quantum statistical approach, the pressure acting on the surface of a homogeneous spherical cavity  sharing center with a sphere of the same material is computed in \cite{hoye2001casimir}.  The medium between both bodies is vacuum, but as the authors mention, their results can be easily generalized considering the same geometry with three different permittivities $\{\varepsilon_1,\varepsilon_2,\varepsilon_M\}$.
In this  case, the coefficient $A_\ell$ defined in \cite{hoye2001casimir}, fulfilling $ {\textup{sgn}\, A_\ell= \textup{sgn}\, \langle p_\textup{int}\rangle}$,  changes to
\begin{equation*}
	A_\ell=\frac{\left[\varepsilon_1(i\kappa)-\varepsilon_{M}(i\kappa)\right] \left[\varepsilon_2(i\kappa)-\varepsilon_{M}(i\kappa)\right] \ell(\ell+1) }{\left[\ell \varepsilon_1(i\kappa)+\varepsilon_{M}(i\kappa) (\ell+1) \right] \left[\ell \varepsilon_{M}(i\kappa)+ \varepsilon_2(i\kappa)(\ell+1)\right]},
\end{equation*}
so the sign of pressure acting on the surface of the  sphere satisfies
Eq.~\eqref{eq:SignPint}.
\\
\section{Repulsive pressure and DLP configuration}\label{sec:TotalPressure}
 {In current experimental setups, such as those mentioned in Sec.~\ref{sec:I},   the interaction force between at least two bodies is measured \cite{garrett2018measurement}.
	This quantity arises from the dependence of the Casimir energy on the distance between bodies, which excludes self-energy contributions. 
	 Indeed, it is not clear if the latter are observationally well-defined \cite{hoye2001casimir}.
	 In the preceding section we have studied the pressure  due to the interaction term of the energy, which might also be the relevant one in this context \cite{hoye2001casimir,brevik2005casimir}. Indeed, we will recover the DLP result for the interaction force from a limiting case of 	$\textup{sgn}\langle p_\textup{int}\rangle$ in Eq.~\eqref{eq:SignPint}. However, there is no straightforward way to measure this pressure in general geometries and effects of different origin, such as hydrodynamic forces in liquid dielectrics, should be taken into account. {In any case, we will determine the sign of the total pressure for certain configurations described by permittivity $\varepsilon$ and permeability ${\mu}$ functions.} The Casimir energy is written as
\begin{equation}
	E_C= E_1+E_2+ E_\text{int}.
\end{equation}
As we have described, the Casimir force between bodies with disjoint support is free of divergences. This is based on the local nature of the heat kernel coefficients, which determine the
divergent part of the vacuum energy \cite{bordag2009advances}. However, the ultraviolet divergences contained in the self-energies can not always be satisfactorily removed. For homogeneous magnetodielectric spheres, although several regularization procedures have been proposed \cite{bordag2009advances,milton2020self}, to our knowledge the Casimir energy {can  only be  defined in two special cases}: when the speed of light is
identical both inside and outside the sphere and in the dilute approximation \cite{milton2001casimir}. In the latter, when the surrounding medium is vacuum, the renormalized self-energy is
\begin{equation}\label{eq:E1}
	E_1^\text{ren}=\dfrac{23}{1536\pi r_1}\left(\varepsilon_{1}-1\right)^2+O\left(\varepsilon_{1}-1\right)^3.
\end{equation}
In addition, the actual computation of the interaction energy could be simplified. Assuming $\varepsilon_i \simeq 1$, we can expand $\log(\mathbb{I}-s\, \mathbb{M})$ in Eq.~\eqref{eq:EnM} as
\begin{equation}
	\text{Tr} \log(\mathbb{I}-s\,\mathbb{M})=-\sum_{j=1}^{\infty}\frac{\text{Tr}(	\mathbb{I}-s\, \mathbb{M})^j}{j}
\end{equation}
and take only the leading terms. Also, the operator $\mathbb{T}_{i}=\mathbb{V}_{i} /({\mathbb{I} + \mathbb{G}^{M} \mathbb{V}_{i}})$ can  be expanded in powers of $\varepsilon_i-1$ \cite{kenneth2008casimir}.

Using the principle of virtual work, noting that
the self-energy of the cavity $E_2$ is independent of $r_1$, we define the total pressure on the sphere as
\begin{equation}\label{eq:TotalPressure}
	\langle p \rangle\equiv -\dfrac{1}{4\pi r_{1}^{2}}\left(\frac{\partial  E^\text{ren}_1}{\partial  r_{1}}+\frac{\partial  E_\text{int}}{\partial  r_{1}}\right).
\end{equation}
We then consider the configuration shown in Fig.~\ref{fig:conf}(a) with dielectric response functions such that $\varepsilon_1$ is close to one and $\varepsilon_2(i\kappa, \textbf{x})> \varepsilon_1$ for the whole cavity.
From the renormalized energy \eqref{eq:E1} it is clear that the self-pressure is repulsive, i.e., it tends to expand the sphere. In addition, from Eq.~\eqref{eq:SignPint}, we know that $\varepsilon_{2}>\varepsilon_{1}>\varepsilon_{M}=1$ also results in a positive interaction pressure. Consequently,  we obtain $\langle p \rangle>0$.

It should be mentioned that the tendency to expand the sphere would be described as an attractive  interaction force  as outlined in Sec.~\ref{sec:I}. We can see it explicitly recovering the DLP result from Eq.~\eqref{eq:SignPint}. Firstly, the planar geometry is reached if we assume a spherical cavity with inner radius $r_2$ and take the limits $r_i\to\infty$, being $d\equiv r_2-r_1$ constant \cite{cavero2020casimir}. The interaction force per unit area $F_\text{int}$ may be defined analogously to the pressure in Eq.~\eqref{eq:MeanPress}, using now  variations of the distance between bodies. Noting that $\partial_{r_1}E_\text{int}=-\partial_{d}E_\text{int}$, from Eq.~\eqref{eq:SignPint} we finally obtain the DLP result \eqref{eq:LifshitzRel}
\begin{equation}\label{eq:SignFint}
	\textup{sgn}\,F_\textup{int} = -\textup{sgn}\left[(\varepsilon_{1}-\varepsilon_{M}\right)(\varepsilon_{2}-\varepsilon_{M})].
\end{equation}
As we have mentioned in Sec.~\ref{sec:I}, this gives rise to repulsive forces if  $\varepsilon_{2}>\varepsilon_{M}>\varepsilon_{1}$,  and to attractive ones if $\varepsilon_{2}>\varepsilon_{1}>~ \varepsilon_{M}$. Both cases have been confirmed experimentally \cite{munday2009measured}. Furthermore, we have proved that the DLP result can be extended to  inhomogeneous slabs as long as Eq.~\eqref{eq:Signs} holds.
\section{Extensions and concluding remarks}\label{sec:IV}
We complete this work presenting some remarks on the main result. In particular, we generalize the configuration considered in Sec.~\ref{sec:III} to magnetodielectric cavities and systems at thermal equilibrium.
\\
{(1) \textit{Magnetodielectric cavity.}} 
The electromagnetic Calogero equation \eqref{eq:Caloguero} requires the homogeneous background and the scattering object to be nonmagnetic. However, in order to determine the sign of $\langle p_\textup{int}\rangle$ we have only assumed a well-defined sign of  $\mathbb{T}_2(i\kappa)$, i.e., $\langle p_\textup{int}\rangle=s_1 s_2$.  Therefore, we can consider a cavity described by functions $\varepsilon_2$ and $\mu_2$, such that
\begin{equation*}
	s_{2}=\pm 1 \quad \text{if} \  \varepsilon_{2}(i\kappa, \textbf{x}) \gtrless \varepsilon_{M}(i \kappa)\ \text{and}\ \mu_{2}(i \kappa, \textbf{x})~\lesseqgtr~\mu_{M}(i \kappa),
\end{equation*}
for the whole body \cite{rahi2009scattering}. 
{In addition, since $E_1$ is independent of the second object, the total pressure defined in Eq.~\eqref{eq:TotalPressure} is positive if $s_2=1$ and $\varepsilon_2(\omega, \textbf{x})> \varepsilon_1$.}
\\
{(2)\textit{ Finite} $T$.}
The extension of results \eqref{eq:SignEn} and \eqref{eq:SignPint} to quantum systems at thermal equilibrium follows from the Matsubara formulation. The free energy $F_\text{int}$  satisfies \cite{milton2001casimir}
\begin{equation}\label{eq:PFint}
	\langle p_\text{int} \rangle =-\frac{1}{4\pi r_{1}^{2}}\,\frac{\partial  F_\text{int}}{\partial  r_{1}},
\end{equation}
and we can compute it  replacing the integral in $E_\textup{int}$ by a sum over the Matsubara frequencies $\kappa_{n}~=~2\pi k_\text{B} {n} T$, where the zero mode is weighted by $1/2$  and the temperature enters as a multiplicative factor \cite{bordag2009advances}:
\begin{equation}
	F_\text{int}= k_B T \sum_{n=0}^{\infty}{}^\prime \ \text{Tr} \log\left[\mathbb{I}-s\, \mathbb{M}(i \kappa_{n})\right].
\end{equation} 
With this, results \eqref{eq:SignEn} and \eqref{eq:SignPint} can be reproduced with minor changes. We simply notice that we have treated each  frequency separately and that  $\partial_{r_{1}}({s_1\mathbb{T}_1})~>~0$, which refers only to electromagnetic scattering, also applies. 
In the dilute approximation, the free energy of the sphere at low temperature is \cite{milton2001casimir} 
$$
F_1^\text{ren}\simeq \left(\varepsilon_{1}-1\right)^2\left[\dfrac{23}{1536\pi r_1}+\dfrac{7}{270}(\pi r_1)^3T^4\right],
$$
where the additional term  counteracts the zero temperature repulsion. Nevertheless, if $T$ is small enough we still obtain a total positive pressure.
\\
{(3) \textit{Energy, force and stable levitation.}}
The Casimir force can switch from attractive to repulsive when the sign of the energy changes. This has been proved in planar geometries for a scalar field satisfying a four parameter  family of  boundary conditions  at the plates \cite{asorey2013attractive} and for inhomogeneous dielectrics slabs, being the the sign of the force given by Eq.~\eqref{eq:SignFint} and the sign of the energy  by Eq.~\eqref{eq:SignEn}.
The same holds for  mirror symmetric objects: the force between them is attractive being the interaction energy  always negative \cite{kenneth2006opposites}. 
In addition, the condition determining unstable  levitation based on Casimir-Lifshitz forces is also governed by the sign of the energy \cite{rahi2010constraints}. In particular,   $s>0$ implies unstable levitation, being 
$
s=-\text{sgn}\, E_\text{int}
$
as we have proved.
Similarly, in the present case, 
$$\textup{sgn}\, E_\textup{int} = -\textup{sgn} \,\langle p_\text{int} \rangle.$$
Indeed, stable levitation is possible if, and only if, the pressure is negative.
However, there are nontrivial configurations where attractive and repulsive forces are found for constant values of $\text{sgn}\, E_\textup{int}$ \cite{levin2010casimir,abrantes2018repulsive}.
\\
{(4) \textit{Exterior configuration.}} For both expansions of the operator ${\mathbb{T}_1}\mathbb{G}^{M}_{12}\mathbb{{T}}_2\mathbb{G}^{M}_{21}$ in an exterior or a cavity configuration, \eqref{eq:ExtConf} and \eqref{eq:IntConf}, the scattering amplitude $\langle\textbf{E}_{i}^\text{rg}, {\mathbb{T}_1} \textbf{E}_{a}^\text{rg}\rangle$ characterizes the first body. As we have already discussed, the scattering amplitudes are only different for the second body. Then, the result on the sign of  the pressure  holds for a exterior configuration, being the net force acting on the sphere in general  nonzero.
\\
\subsection*{Conclusions}
We are able to determine the tendency of the interaction pressure to expand or contract the sphere. The result does not depend either on the geometry of the cavity or on the matter distribution inside the sphere, as long as the assumption \eqref{eq:Signs} holds.  Since the proof applies to each frequency independently, the extension to systems at thermal equilibrium is almost immediate.
We find that the sign of the pressure changes with the sign of $(\varepsilon_1-\varepsilon_{M})(\varepsilon_2-\varepsilon_{M})$.
This behavior was first found by DLP  in \cite{dzyaloshinskii1961general}, where they extended
Casimir's
formulation for ideal metal plates in vacuum to dielectric
materials.
Indeed,  the same pattern arises when determining  stable levitation based on Casimir-Lifshitz forces using the scattering approach \cite{rahi2010constraints}.   
Within this approach we have obtained the DLP result, and its extension to inhomogeneous slabs, as a limiting case. Indeed, spatial dispersion could have been included with nonlocal potentials $\mathbb{V}_{i}(\omega, \textbf{x},\textbf{x}')$ \cite{rahi2009scattering}, being this outside the application region of Lifshitz theory \cite{klimchitskaya2007comment}.
The self-energy contribution to the pressure can be added for configurations in which the ultraviolet divergences can be satisfactorily removed.
We have illustrated this fact obtaining a total positive pressure for a dilute dielectric ball enclosed within an arbitrarily shaped magnetodielectric cavity.

\begin{acknowledgements}
I am grateful to I. Cavero\,-Pel\'aez,  A. Romaniega, L. M. Nieto,  and J. M. Mu\~noz-Casta\~neda for the useful suggestions. This work was  supported by the FPU fellowship program (FPU17/01475) and the  Junta de Castilla y Le\'on and FEDER projects (BU229\-P18 and VA137G18). 
\end{acknowledgements}


\begin{thebibliography}{10}

	\bibitem{bordag2009advances}
	M.~Bordag, G.~L. Klimchitskaya, U.~Mohideen, V.~M. Mostepanenko,
	\newblock {\em Advances in the Casimir effect}
	\newblock (Oxford
	University Press, Oxford, 2009)
	
	\bibitem{milton2001casimir}
	K.~A. Milton,
	\newblock {\em The Casimir effect: physical manifestations of zero-point
		energy}
	\newblock (World Scientific, Singapore, 2001)
	
	\bibitem{kenneth2006opposites}
	O.~Kenneth, I.~Klich,
	\newblock {Phys. Rev. Lett.} \textbf{97}, 160401 (2006)
	
	\bibitem{bachas2007comment}
	C.~P. Bachas,
	\newblock {J. Phys. A} \textbf{40}, 9089 (2007)
	
	\bibitem{buks2001stiction}
	E.~Buks, M.~L. Roukes,
	\newblock {Phys. Rev. B} \textbf{63}, 033402 (2001)
	
	\bibitem{munday2010repulsive}
	J.~N. Munday, F.~Capasso,
	\newblock {Int. J. Mod. Phys. A} \textbf{25}, 2252 (2010)
	
	\bibitem{boyer1974van}
	T.~H. Boyer,
	\newblock {Phys. Rev. A} \textbf{9}, 2078 (1974)
	
	\bibitem{rosa2008casimir}
	F.~S.~S. Rosa, D.~A.~R. Dalvit, P.~W. Milonni,
	\newblock {Phys. Rev. Lett.} \textbf{100}, 183602  (2008)
	
	\bibitem{zhao2009repulsive}
	R.~Zhao, J.~Zhou, Th. Koschny, E.~N. Economou, C.~M. Soukoulis,
	\newblock {Phys. Rev. Lett.} \textbf{103}, 103602 (2009)
	
	\bibitem{grushin2011tunable}
	A.~G. Grushin, A.~Cortijo,
	\newblock {Phys. Rev. Lett.} \textbf{106}, 020403  (2011)
	\bibitem{rodriguez2014repulsive}
	P. Rodriguez-Lopez, A. G. Grushin,    Phys. Rev. Lett. \textbf{112}, 056804 (2014)
	\bibitem{levin2010casimir}
	M.~Levin, A.~P. McCauley, A.~W. Rodriguez,
	M.~T.~H. Reid, S.~G.	Johnson,
	\newblock {Phys. Rev. Lett.} \textbf{105}, 090403 (2010)
	
	\bibitem{abrantes2018repulsive}
	P.~P. Abrantes, Y.~Fran{\c{c}}a, F.~S.~S. da~Rosa, C.~Farina, R. de~Melo~e~Souza,
	\newblock {Phys. Rev. A} \textbf{98}, 012511 (2018)
	
	\bibitem{dzyaloshinskii1961general}
	I.~E. Dzyaloshinskii, E.~M. Lifshitz, L.~P. Pitaevskii,
	\newblock {Adv. Phys.} \textbf{10}, 165 (1961)
	
	\bibitem{munday2009measured}
	J.~N. Munday, F.~Capasso, V.~A. Parsegian,
	\newblock {Nature} \textbf{457}, 170 (2009)
	
	\bibitem{venkataram2020fundamental}
	P.~S. Venkataram, S.~Molesky, P.~Chao, A.~W. Rodriguez,
	\newblock {Phys. Rev. A} \textbf{101}, 052115 (2020)
	
	\bibitem{capasso2007casimir}
	F. Capasso, J. N. Munday, D. Iannuzzi, H. B. Chan, IEEE J.  Quantum Electron.  \textbf{13},  400 (2007)
	
	\bibitem{rahi2010constraints}
	S.~J. Rahi, M.~Kardar, T.~Emig,
	\newblock {Phys. Rev. Lett.} \textbf{105}, 070404 (2010)
	
	\bibitem{jiang2019chiral}
	Q.~D. Jiang, F.~Wilczek,
	\newblock {Phys. Rev. B} \textbf{99}, 125403 (2019)
	
	\bibitem{marachevsky2001casimir}
	V.~N. Marachevsky,
	\newblock {Phys. Scr.} \textbf{64}, 205 (2001)
	
	\bibitem{hoye2001casimir}
	J.~S. H{\o}ye, I.~Brevik, J.~B. Aarseth,
	\newblock {Phys. Rev. E} \textbf{63}, 051101 (2001)
	
	\bibitem{brevik2002casimir}
	I~Brevik, J.~B. Aarseth, J.~S. H{\o}ye,
	\newblock {Phys. Rev. E} \textbf{66}, 026119 (2002)
	
	\bibitem{brevik2005casimir}
	I.~Brevik, E.~K. Dahl, G.~O. Myhr,
	\newblock {J. Phys. A Math. Gen.} \textbf{38}, L49 (2005)
	
	\bibitem{dalvit2006exact}
	D.~A.~R. Dalvit, F.~C. Lombardo, F.~D. Mazzitelli, R.~Onofrio,
	\newblock {Phys. Rev. A} \textbf{74}, 020101(R) (2006)
	
	\bibitem{marachevsky2007casimir}
	V.~N. Marachevsky,
	\newblock {Phys. Rev. D} \textbf{75}, 085019 (2007)
	
	\bibitem{zaheer2010casimir}
	S.~Zaheer, S.~J. Rahi, T.~Emig, R.~L. Jaffe,
	\newblock {Phys. Rev. A} \textbf{82}, 052507 (2010)
	
	\bibitem{teo2010casimir}
	L.~P. Teo,
	\newblock {Phys. Rev. D} \textbf{82}, 085009 (2010)
	
	\bibitem{rahi2010stable}
	S.~J. Rahi, S.~Zaheer,
	\newblock {Phys. Rev. Lett.} \textbf{104}, 070405 (2010)
	
	\bibitem{parashar2017electromagnetic}
	P.~Parashar, K.~A. Milton, K.~V. Shajesh, I.~Brevik,
	\newblock {Phys. Rev. D} \textbf{96}, 085010 (2017)
	
	\bibitem{kenneth2008casimir}
	O.~Kenneth, I.~Klich,
	\newblock {Phys. Rev. B} \textbf{78}, 014103 (2008)
	
	\bibitem{rahi2009scattering}
	S.~J. Rahi, T.~Emig, N.~Graham, R.~L. Jaffe, M.~Kardar,
	\newblock {Phys. Rev. D} \textbf{80}, 085021 (2009)
	
	\bibitem{newton2013scattering}
	R.~G. Newton,
	\newblock {\em Scattering theory of waves and particles}
	\newblock (Dover, Mineola, New York, 2002)
	
	\bibitem{hanson2013operator}
	G.~W. Hanson, A.~B. Yakovlev,
	\newblock {\em Operator theory for electromagnetics}
	\newblock (Springer, New York, 2002)
	
	\bibitem{sun2019invariant}
	B.~Sun, L.~Bi, P.~Yang, M.~Kahnert, G.~Kattawar,
	\newblock {\em Invariant Imbedding T-matrix method for light scattering by
		nonspherical and inhomogeneous particles}
	\newblock (Elsevier, Amsterdam, 2019)
	
	\bibitem{teo2012mode}
	L.~P. Teo,
	\newblock {Int. J. Mod. Phys. A} \textbf{27}, 1230021 (2012)
	
	\bibitem{simon1977notes}
	B.~Simon,
	\newblock {Adv. Math.} \textbf{24}, 244 (1977)
	
	\bibitem{barton2004casimir}
	G.~Barton,
	\newblock {J. Phys. A Math. Gen.} \textbf{37}, 3725 (2004)
	
	\bibitem{li2019casimir}
	Y.~Li, K.~A. Milton, X.~Guo, G.~Kennedy, S.~A. Fulling,
	
	
	\bibitem{johnson1988invariant}
	B.~R. Johnson,
	\newblock {Appl. Opt.} \textbf{27}, 4861 (1988)
	
	\bibitem{forrow2012variable}
	A.~Forrow, N.~Graham,
	\newblock {Phys. Rev. A} \textbf{86}, 062715 (2012)
	
	\bibitem{calogero1967variable}
	F.~Calogero,
	\newblock {\em Variable Phase Approach to Potential Scattering}
	\newblock (Academic, New York, 1967)
	
	\bibitem{johnson1999exact}
	B.~R. Johnson,
	\newblock {J. Opt. Soc. Am. A} \textbf{16}, 845 (1999)
	
	\bibitem{toni2014advances}
	B.~Toni,
	\newblock {\em Advances in Interdisciplinary Mathematical Research} 
	\newblock (Springer, New York, 2014),
	\newblock Chapter 3.
	
	\bibitem{feynman1939forces}
	R.~P. Feynman,
	\newblock {Phys. Rev.} \textbf{56}, 340 (1939)
	
	
	\bibitem{garrett2018measurement}
	J. L. Garrett,  D. A. T. Somers, J. N. Munday,
	\newblock {Phys. Rev. Lett.} \textbf{120}, 040401 (2018)
	
	\bibitem{milton2020self}
	K.~A. Milton,  P. Parashar,   I. Brevik,  G. Kennedy,
	\newblock {Ann. Phys.} \textbf{412}, 168008 (2020)
	
	\bibitem{cavero2020casimir}
	I. Cavero\,-Pel\'aez,  J. M. Mu\~noz-Casta\~neda, C. Romaniega,
	\newblock {Phys. Rev. D} (to be published) [accepted for publication]	arXiv:2009.03785.
	
	\bibitem{asorey2013attractive}
	M.~Asorey, J.~M. Munoz-Castaneda,
	\newblock {Nucl. Phys. B} \textbf{874}, 852 (2013)
	
	\bibitem{klimchitskaya2007comment}  G. L. Klimchitskaya, V. M. Mostepanenko,
	\newblock {Phys. Rev. B}  \textbf{75}(3), 036101 (2007)
	
\end{thebibliography}
\end{document}